\IfFileExists{ws-ijgmmp.cls}{\documentclass{ws-ijgmmp}\def\usingWS{1}}{\documentclass[11pt]{article}}
\usepackage{amsmath,amssymb,graphicx}
\ifdefined\usingWS\else
\usepackage[margin=2.6cm]{geometry}
\usepackage[T1]{fontenc}
\makeatletter
\def\@authlist{}
\renewcommand\author[1]{\g@addto@macro\@authlist{{\scshape #1}\par}}
\newcommand\address[1]{\g@addto@macro\@authlist{{\small\itshape #1}\par\medskip}}
\newcommand\email[1]{\g@addto@macro\@authlist{{\small\ttfamily #1}\par\medskip}}
\newcommand\keywords[1]{\par\medskip\noindent{\small\textit{Keywords}: #1}}
\newcommand\ccode[1]{\par\noindent{\small #1}}
\newcommand\catchline[5]{}
\renewcommand\maketitle{\begin{center}{\Large\bfseries\@title\par}\bigskip\@authlist\end{center}}
\makeatother
\fi

\begin{document}
\catchline{}{}{}{}{}

\markboth{A. ? et al.}{Acoustic scattering and radiation force on a global-monopole defect}

\title{Acoustic scattering and radiation force on the core of a\\ global-monopole defect}

\author{Átila Pereira Ribeiro}
\address{Instituto Federal de Alagoas (IFAL),
Campus Palmeira dos Índios,\\
Av. Alagoas, s/n, Palmeira de Fora,
57608-180 Palmeira dos Índios, AL, Brazil\\
atila.ribeiro@ifal.edu.br}

\author{Alexandre M. de M. Carvalho}
\address{Instituto de F\'{\i}sica,
Universidade Federal de Alagoas,\\
Campus A.~C.~Sim\~oes,
57072-970 Macei\'o, AL, Brazil\\
alexandre@fis.ufal.br}

\author{Claudio Furtado}
\address{Departamento de F\'{\i}sica, CCEN,
Universidade Federal da Para\'{\i}ba,\\
58051-970 Jo\~ao Pessoa, PB, Brazil\\
furtado@fisica.ufpb.br}

\maketitle

\begin{abstract}
We investigate the scattering of scalar acoustic waves by the rigid core of a point defect described by the global-monopole geometry, \(dl^2=dr^2+\alpha^2r^2d\Omega^2\), allowing for both a deficit (\(\alpha<1\)) and an excess (\(\alpha>1\)) of solid angle. We show that this geometry can be realized exactly in a fluid with constant radial and tangential mass densities, where \(\alpha^2=\rho_t/\rho_r\). Because the metric is not asymptotically flat, a global plane wave cannot be defined, and we formulate the scattering problem in terms of partial-wave phase shifts. These separate exactly into a topological contribution, \(\frac{\pi}{2}(\ell-s_\ell)\), and a core contribution determined by spherical Bessel functions of noninteger order \(s_\ell\). We derive a closed-form expression for the topological amplitude, supplemented by an absolutely convergent remainder. The amplitude is singular on the cone \(\theta_0=\pi|1-1/\alpha|\), corresponding to the universal deflection angle of acoustic geodesics. The core amplitude obeys a generalized Rayleigh law, with a dipole exponent \(\sqrt{8+\alpha^2}/\alpha\). In the absence of translational isometries, the radiation force must be evaluated directly on the core surface. We derive a partial-wave expression for this force and show that its leading low-frequency contribution is linear in the topological phase. For every \(\alpha>1/\sqrt{6}\), except \(\alpha=1\), the force is negative within a low-frequency interval \(ka<ka_c(\alpha)\), pulling the core toward the source. For \(\alpha>\alpha_*\simeq1.372\), the normalized force diverges as \(ka\to0\).
\end{abstract}



\section{Introduction}
\label{sec:intro}

Condensed-matter systems have long served as laboratories for exploring
gravitational and cosmological phenomena~\cite{MoraesBJP,BLV}.
Unruh's observation that sound propagation in a moving fluid can be
described in terms of an effective Lorentzian geometry~\cite{Unruh81}
paved the way for analog models in classical fluids~\cite{Visser98},
Bose--Einstein condensates~\cite{Garay}, superfluid helium~\cite{Volovik},
and optical media. In a related approach, the geometric theory of
defects~\cite{Kroner,Bilby,Kleinert,KatanaevVolovich,Katanaev}
describes elastic solids containing disclinations and dislocations
in terms of Riemann--Cartan geometry, with curvature and torsion
associated with the defects. This framework has been used to
investigate holonomies and Aharonov--Bohm-type phases arising from
the scattering of phonons and quasiparticles by defects in both
superfluids~\cite{FurtadoAB,CarvalhoEPL} and elastic
media~\cite{FurtadoElastic}.

Spherically symmetric point defects, such as hedgehogs in ordered
media or defects formed by removing or inserting solid angle in a
solid, are the condensed-matter counterparts of the global monopole
introduced by Barriola and Vilenkin~\cite{Barriola}.
Their effective geometry exhibits a deficit or excess of solid angle
that persists at infinity, together with a scalar curvature that
falls off as $r^{-2}$. Mazur and Papavassiliou~\cite{Mazur,Puszkarz}
investigated field scattering by a pointlike global monopole,
building on earlier studies of scattering in conical
geometries~\cite{DeserJackiw,tHooft}.

However, two questions of particular relevance to acoustics remain
open. The first concerns how long-range topological scattering
combines with scattering by a finite-sized defect core. The second
concerns the mechanical action of the acoustic wave on that core,
namely the acoustic radiation force~\cite{King,Gorkov,Westervelt,SilvaJASA}.
The latter is particularly interesting because negative radiation
forces, commonly associated with acoustic tractor beams, are typically achieved by shaping the incident
beam~\cite{Marston2006,ZhangMarston,Demore} rather than by modifying the surrounding medium.

In this paper, we consider a rigid core of radius $a$ at the center
of a global-monopole defect in an acoustic medium. We first show,
in Sec.~2, that the global-monopole geometry can be realized exactly
in a fluid with constant radial and tangential mass densities,
where $\alpha^2=\rho_t/\rho_r$. This allows, in principle, for both
$\alpha<1$ and $\alpha>1$. In Sec.~3, we formulate the scattering
problem in terms of partial-wave phase shifts and separate the
scattering amplitude into topological and core contributions.
We obtain the topological amplitude in closed form, up to an
absolutely convergent remainder, without expanding in the defect
strength. Its singularity at
$\theta_0=\pi|1-1/\alpha|$ coincides with the universal deflection
angle of acoustic rays. The core contribution follows a generalized
Rayleigh law and, for $\alpha>1$, is dominated by the dipole term,
with an intensity proportional to $(ka)^{4s_1}$ rather than
$(ka)^4$.

In Sec.~4, we investigate the acoustic radiation force on the core.
Because the medium has no translational isometries, the force must
be evaluated directly on the core surface. We derive a partial-wave
expression for this force and show that, for every
$\alpha>1/\sqrt{6}$ except $\alpha=1$, the core experiences a
pulling force within a low-frequency interval
$ka<ka_c(\alpha)$. Finally, Sec.~5 summarizes our main findings.


\section{Effective geometry and its realizations}
\label{sec:geometry}

\subsection{The global-monopole metric}

The global monopole of Barriola and Vilenkin~\cite{Barriola} is produced by the
spontaneous breaking of a global $O(3)$ symmetry. Away from its core the
spacetime is well approximated by
$ds^2=dt^2-dr^2-\alpha^2r^2(d\theta^2+\sin^2\theta\,d\varphi^2)$, where
$\alpha^2=1-8\pi G\eta^2$ and $\eta$ sets the symmetry-breaking scale. In
cosmology $0<\alpha<1$. In the condensed-matter and analog settings we have in
mind, however, $\alpha$ is a material parameter, and a deficit ($\alpha<1$) and
an excess ($\alpha>1$) of solid angle are equally admissible. What matters for
wave propagation is the spatial metric
\begin{equation}
dl^2 = dr^2 + \alpha^2 r^2\left(d\theta^2+\sin^2\theta\, d\varphi^2\right).
\label{eq:metric}
\end{equation}
A short calculation gives $R_{rr}=0$ and
$R_{\theta\theta}=R_{\varphi\varphi}/\sin^2\theta=1-\alpha^2$ for the Ricci
tensor, and hence the scalar curvature
\begin{equation}
\mathcal R = \frac{2\left(1-\alpha^2\right)}{\alpha^2 r^2}.
\label{eq:curvature}
\end{equation}
The medium is therefore not locally flat, which sets it apart from the cone
associated with a cosmic string or a wedge disclination. Its isometry group is
$SO(3)$; for $\alpha\neq1$ there are no translational Killing vectors, a point
that will be decisive for the momentum balance in Sec.~\ref{sec:force}.

\subsection{Wave equation and separation of variables}

We describe the acoustic field by a velocity potential $\phi$, with pressure
$p=i\omega\rho_0\phi$. It satisfies the Helmholtz equation
$(\Delta_{\rm LB}+k^2)\phi=0$, where $\Delta_{\rm LB}$ is the Laplace--Beltrami
operator of (\ref{eq:metric}):
\begin{equation}
\frac{1}{r^2}\frac{\partial}{\partial r}\left(r^2\frac{\partial\phi}{\partial r}\right)
+\frac{1}{\alpha^2r^2}\,\hat L\phi+k^2\phi=0,\qquad
\hat L=\frac{1}{\sin\theta}\frac{\partial}{\partial\theta}\left(\sin\theta\frac{\partial}{\partial\theta}\right)+\frac{1}{\sin^2\theta}\frac{\partial^2}{\partial\varphi^2}.
\label{eq:helmholtz}
\end{equation}
Writing $\phi=R_\ell(r)Y_\ell^m(\theta,\varphi)$, with
$\hat LY_\ell^m=-\ell(\ell+1)Y_\ell^m$, we obtain the radial equation
\begin{equation}
R_\ell''+\frac{2}{r}R_\ell'+\left[k^2-\frac{\ell(\ell+1)}{\alpha^2r^2}\right]R_\ell=0 .
\label{eq:radial}
\end{equation}
It is solved by spherical Bessel functions $j_{s_\ell}(kr)$ and
$y_{s_\ell}(kr)$ of non-integer order, fixed by
$s_\ell(s_\ell+1)=\ell(\ell+1)/\alpha^2$:
\begin{equation}
s_\ell=\frac{\sqrt{4\ell(\ell+1)+\alpha^2}-\alpha}{2\alpha}
=\nu_\ell-\frac12,\qquad
\nu_\ell=\frac1\alpha\sqrt{\left(\ell+\tfrac12\right)^2-a_d},\qquad
a_d\equiv\frac{1-\alpha^2}{4}.
\label{eq:order}
\end{equation}
In terms of ordinary Bessel functions, $R_\ell=(kr)^{-1/2}\{J_{\nu_\ell},Y_{\nu_\ell}\}(kr)$.
The angular dependence is the same as in flat space; the defect acts only
through the shifted orders $s_\ell$. Note that $s_0=0$ for any $\alpha$, so the
$s$-wave does not feel the defect at all.

\subsection{Physical realizations}
\label{sec:realisation}

The first and most natural setting is that of point defects in elastic media and
ordered phases. In the geometric theory of
defects~\cite{Kroner,Bilby,Kleinert,KatanaevVolovich,Katanaev}, a solid with
defects is a Riemann--Cartan manifold, disclinations being sources of curvature
and dislocations sources of torsion. A spherically symmetric point defect,
formed by removing or inserting solid angle, is then described outside its
core by a metric of the form (\ref{eq:metric}). Closely related hedgehog
defects occur in superfluid $^3$He and in nematic liquid crystals, where they
give rise to lensing of the global-monopole
type~\cite{Volovik,SatiroMoraes,MoraesBJP}. In a solid, Eq.~(\ref{eq:helmholtz})
describes the longitudinal (compressional) mode once conversion into shear
waves is neglected. This scalar approximation was also used in earlier work on
phonons scattered by disclinations~\cite{FurtadoElastic}.

For fluids, a stronger statement can be made: the geometry can be realized
exactly. Take a fluid whose mass density is radially anisotropic, with value
$\rho_r$ for motion along $\hat{\mathbf r}$ and $\rho_t$ for tangential motion,
and whose bulk modulus is $\kappa$. The pressure then obeys
$\nabla\cdot(\hat\rho^{-1}\nabla p)+\omega^2p/\kappa=0$, which in spherical
coordinates reads
\begin{equation}
\frac{1}{\rho_r}\frac{1}{r^2}\partial_r\left(r^2\partial_rp\right)
+\frac{1}{\rho_t}\frac{1}{r^2}\hat Lp+\frac{\omega^2}{\kappa}p=0 .
\end{equation}
After multiplication by $\rho_r$ this is precisely Eq.~(\ref{eq:helmholtz}), with
\begin{equation}
\alpha^2=\frac{\rho_t}{\rho_r},\qquad k=\omega\sqrt{\rho_r/\kappa}.
\label{eq:metafluid}
\end{equation}
A metafluid with spherical anisotropy and constant spherical components
$\rho_r$, $\rho_t$ and $\kappa$ thus reproduces the global-monopole geometry
without any radial grading of its parameters. Its Cartesian density tensor does
depend on direction, and it is singular at the origin; this is where the core
sits. The construction is a particular case of the transformation-acoustics
correspondence between the Laplace--Beltrami operator and fluids with
anisotropic density~\cite{CummerSchurig,ChenChan,Norris}.

Choosing $\rho_r=\rho_0$, $\rho_t=\alpha^2\rho_0$ and $\kappa=\kappa_0$ keeps the
radial impedance equal to that of the host fluid. Radially anisotropic acoustic
media have already been designed from concentric layers of two
fluids~\cite{TorrentPRL}. For layers normal to $\hat{\mathbf r}$, the effective
radial density is the arithmetic mean of the constituent densities and the
tangential density is their harmonic mean, so layering can only give
$\rho_t\le\rho_r$, that is, $\alpha\le1$. Reaching $\alpha>1$ requires a
microstructure that hinders tangential motion, such as radially oriented
plate-like inclusions. In either case a rigid inclusion at the center plays the
part of the defect core. A metafluid ball of finite radius $R\gg a$ would add an
outer interface and a scattering problem of its own, which we leave aside;
Secs.~\ref{sec:scattering} and~\ref{sec:force} treat the medium
(\ref{eq:metric}) as filling all space.


\section{Scattering by a rigid core in the monopole medium}
\label{sec:scattering}

Consider now a monochromatic wave $\phi(\mathbf r)e^{-i\omega t}$, with
$k=\omega/c_0$, propagating in the medium (\ref{eq:metric}) and scattered by the
defect core. We model the core as an immovable rigid sphere of radius $a$
centered at $r=0$. Besides being the scatterer, it removes the singular point
at the origin. Since the deficit or excess of solid angle extends to infinity,
the metric is not asymptotically flat and no plane wave can be defined
globally; indeed, $e^{ikr\cos\theta}$ is not a solution of
Eq.~(\ref{eq:helmholtz}) when $\alpha\neq1$, and cannot serve as the incident
field. As for the cone~\cite{DeserJackiw,tHooft} and for the gravitational
global monopole~\cite{Mazur,Puszkarz}, the scattering problem has to be posed in
terms of asymptotic partial waves.

\subsection{Geometric acoustics: universal deflection}
\label{sec:rays}

\begin{figure}[t]
\centering
\includegraphics[width=0.95\linewidth]{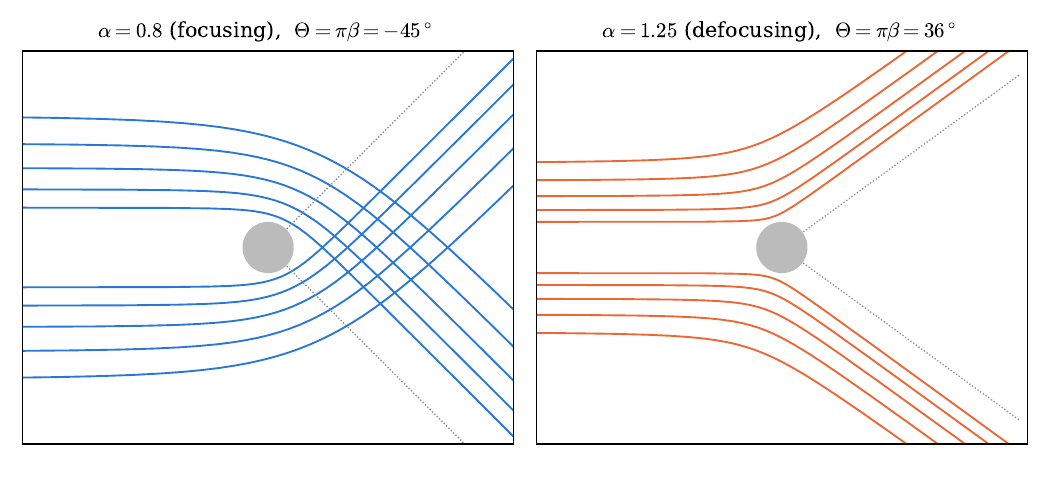}
\caption{Acoustic rays (geodesics of Eq.~(\ref{eq:metric})) in the plane
$\theta=\pi/2$ for $\alpha=0.8$ (left) and $\alpha=1.25$ (right). The gray disk
is the rigid core. All rays that miss the core are deflected by the same angle
$\Theta=\pi(1-1/\alpha)$ (dotted lines).}
\label{fig:rays}
\end{figure}

We first consider the ray limit, $ka\gg1$, in which sound
travels along the geodesics of (\ref{eq:metric}). By spherical symmetry every
ray lies in a plane through the origin, which we may take to be
$\theta=\pi/2$; there $dl^2=dr^2+\alpha^2r^2d\varphi^2$. With
$\tilde\varphi=\alpha\varphi$ this metric becomes locally Euclidean, so the
geodesics are straight lines in $(r,\tilde\varphi)$,
$r=b/\sin\tilde\varphi$ with $0<\tilde\varphi<\pi$ and impact parameter $b$.
Along a complete ray the coordinate angle $\varphi$ therefore advances by
$\pi/\alpha$ rather than $\pi$, and every ray with $b>a$ is deflected by the
same angle,
\begin{equation}
\Theta = \pi\beta,\qquad \beta\equiv 1-\frac1\alpha ,
\label{eq:deflection}
\end{equation}
whatever its impact parameter or frequency. For $\alpha<1$ the rays cross the
symmetry axis and the defect focuses; for $\alpha>1$ they are bent away from
it. Either way, the classical cross section is concentrated on the cone
$\theta=\theta_0$, where $\cos\theta_0=\cos\pi\beta$ and $0\le\theta_0\le\pi$
(Fig.~\ref{fig:rays}); for $\alpha\ge\frac12$ this is simply
$\theta_0=\pi|\beta|$. As we shall see, this ring is where the wave amplitude
becomes singular.

\subsection{Partial waves and phase shifts}
\label{sec:phases}

Outside the core, the partial wave of angular momentum $\ell$ is a combination
of $j_{s_\ell}(kr)$ and $y_{s_\ell}(kr)$, with $s_\ell$ given by
Eq.~(\ref{eq:order}). The rigid boundary condition $\partial_r\phi|_{r=a}=0$
selects
\begin{equation}
R_\ell(r)\propto \cos\eta_\ell\, j_{s_\ell}(kr)-\sin\eta_\ell\, y_{s_\ell}(kr),
\qquad
\tan\eta_\ell=\frac{j'_{s_\ell}(ka)}{y'_{s_\ell}(ka)} .
\label{eq:eta}
\end{equation}
At large distances
$R_\ell\sim\sin\!\left(kr-\tfrac{\pi}{2}s_\ell+\eta_\ell\right)/kr$, and a
comparison with the free wave $\sin(kr-\ell\pi/2)/kr$ shows that the phase shift
is the sum of a topological and a core contribution,
\begin{equation}
\delta_\ell=\delta^{\rm top}_\ell+\eta_\ell,\qquad
\delta^{\rm top}_\ell=\frac\pi2\left(\ell-s_\ell\right)=\frac\pi2\left(\ell+\tfrac12-\nu_\ell\right).
\label{eq:phase}
\end{equation}
The decomposition is exact. Accordingly, the scattering amplitude
\begin{equation}
f(\theta)=\frac{1}{2ik}\sum_{\ell=0}^{\infty}(2\ell+1)\left(e^{2i\delta_\ell}-1\right)P_\ell(\cos\theta)
\label{eq:f}
\end{equation}
splits as $f=f_{\rm top}+f_{\rm core}$, with
\begin{align}
f_{\rm top}(\theta)&=\frac{1}{2ik}\sum_{\ell}(2\ell+1)\left(e^{2i\delta^{\rm top}_\ell}-1\right)P_\ell(\cos\theta),
\label{eq:ftop}\\
f_{\rm core}(\theta)&=\frac{1}{2ik}\sum_{\ell}(2\ell+1)\,e^{2i\delta^{\rm top}_\ell}\left(e^{2i\eta_\ell}-1\right)P_\ell(\cos\theta).
\label{eq:fcore}
\end{align}
The core contribution corresponds to the rigid-sphere amplitude evaluated with the modified orders $s_\ell$ and modified by the topological phase $e^{2i\delta^{\rm top}_\ell}$.
Both familiar limits are contained in Eqs.~(\ref{eq:phase})--(\ref{eq:fcore}).
For $\alpha=1$ we have $s_\ell=\ell$ and $\delta^{\rm top}_\ell=0$, and $f$ is the
classical rigid-sphere amplitude~\cite{MorseIngard}. For $a\to0$, on the other
hand, $\eta_\ell\to0$ and only $f_{\rm top}$ survives, which is the amplitude of
a point-like global monopole~\cite{Mazur}. Particular care is required when choosing the normalization. The coefficients $(2\ell+1)i^\ell$
do not describe a beam of unit intensity in the geometry (\ref{eq:metric}),
because two factors of $\alpha^2$ come in. The first is geometric: a sphere of
radius $r$ has area $4\pi\alpha^2r^2$, so each partial wave brings in $\alpha^2$
times the flux it would carry in flat space. The second comes from the impact
parameter. The turning point of Eq.~(\ref{eq:radial}) places the $\ell$-th wave
at $b_\ell=\nu_\ell/k\simeq\ell/(\alpha k)$, so neighboring partial waves share a
transverse area $2\pi b\,db\simeq2\pi\ell/(\alpha^2k^2)$, smaller by $\alpha^2$
than in flat space. Taken together, unit coefficients correspond to an incident
intensity $I_{\rm inc}=\alpha^4I_0$, where $I_0$ is the intensity of a unit plane
wave in a homogeneous medium. Since the power scattered into $d\Omega$ is
$\alpha^2|f|^2I_0\,d\Omega$, the differential cross section per unit physical
solid angle, $d\Omega_{\rm phys}=\alpha^2d\Omega$, is
\begin{equation}
\frac{d\sigma}{d\Omega_{\rm phys}}=\frac{|f(\theta)|^2}{\alpha^4} .
\label{eq:dcs}
\end{equation}
The total cross section, by contrast, does not exist. Because
$\delta^{\rm top}_\ell\simeq\frac\pi2\beta(\ell+\frac12)$ grows linearly with
$\ell$, the series $\sigma=(4\pi/k^2)\sum(2\ell+1)\sin^2\delta_\ell$ diverges for
every $\alpha\neq1$. The situation is the same as in Aharonov--Bohm and
cosmic-string scattering: $\sigma$ is infinite, and the optical theorem cannot
be invoked to give it a finite value. Physically meaningful quantities are
$d\sigma/d\Omega_{\rm phys}$ away from the forward direction and quantities
evaluated on the core itself, such as the radiation force of
Sec.~\ref{sec:force}.

\subsection{The topological amplitude in closed form}
\label{sec:ftop}

Since its terms do not decay, the series (\ref{eq:ftop}) does not converge in
the ordinary sense. It defines a distribution, which we evaluate as an Abel
limit. Setting $z=\ell+\frac12$ and $t=e^{i\pi\beta}$, Eq.~(\ref{eq:phase}) can
be rewritten without approximation as
\begin{equation}
e^{2i\delta^{\rm top}_\ell}=t^{z}e^{2i\Delta_\ell},\qquad
\Delta_\ell=\frac{\pi}{2\alpha}\left(z-\sqrt{z^2-a_d}\right)=\frac{\pi a_d}{4\alpha z}+O(z^{-3}).
\label{eq:Delta}
\end{equation}
The two generating functions of the Legendre polynomials that we need are
\begin{equation}
h(\theta)=\sum_{\ell}t^{z}P_\ell(\cos\theta)=\frac{t^{1/2}}{\left(1-2t\cos\theta+t^2\right)^{1/2}},\qquad
g(\theta)=\sum_{\ell}(2\ell+1)t^{z}P_\ell(\cos\theta)=\frac{t^{1/2}\left(1-t^2\right)}{\left(1-2t\cos\theta+t^2\right)^{3/2}}.
\end{equation}
With their help we find, for $\theta\neq0$,
\begin{equation}
f_{\rm top}=f^{(0)}+f^{(1)}+\mathcal{F},
\label{eq:ftop_split}
\end{equation}
\begin{equation}
f^{(0)}(\theta)=\frac{g(\theta)}{2ik}=-\frac{\sin\pi\beta}{2\sqrt2\,k}\,\frac{1}{\left(\cos\pi\beta-\cos\theta\right)^{3/2}},\qquad
f^{(1)}(\theta)=\frac{\pi a_d}{2\alpha k}\,h(\theta),
\label{eq:f0f1}
\end{equation}
where the Abel limit fixes the branch through
$\cos\pi\beta\to\cos\pi\beta-i0\sin\pi\beta$ (this only matters inside the ring,
$\theta<\theta_0$), and
\begin{equation}
\mathcal{F}(\theta)=\frac{1}{2ik}\sum_{\ell}(2\ell+1)\,t^{z}\left[e^{2i\Delta_\ell}-1-\frac{i\pi a_d}{2\alpha z}\right]P_\ell(\cos\theta).
\label{eq:remainder}
\end{equation}
The terms of $\mathcal F$ fall off as $z^{-3/2}$, so this remainder converges
absolutely. The leading term $f^{(0)}$ blows up as $|\theta-\theta_0|^{-3/2}$
on the ring $\theta=\theta_0$, which is the wave counterpart of the universal
deflection (\ref{eq:deflection}). Because $f_{\rm top}$ is independent of $a$
and $f_{\rm core}$ is bounded, the singularity is present for any core size and
any frequency. In practice it would be cut off by the finite extent of the
anisotropic region, by absorption, or by the finite width of the incident beam. Although retaining only $f^{(0)}$, may seem sufficient, the first correction is not generally small. At $\alpha=1.25$ and $\theta=2$, for example,
$|f^{(0)}|:|f^{(1)}|:|\mathcal F|\simeq 0.153:0.113:0.031$ in units of $1/k$.
The decomposition (\ref{eq:ftop_split}) needs no expansion in $a_d$ at all: the
closed forms absorb the part of the series that does not converge, and
$\mathcal F$ can be summed numerically to whatever accuracy is required.

\subsection{The core amplitude and a generalized Rayleigh law}
\label{sec:rayleigh}

The core amplitude (\ref{eq:fcore}) is much better behaved, since $\eta_\ell$
decays faster than any power once $\ell$ exceeds $ka$. Its low-frequency limit
follows from the small-argument forms of $j_s$ and $y_s$ for non-integer
$s>0$, which give
\begin{equation}
\tan\eta_\ell\simeq\frac{\pi\,s_\ell}{(s_\ell+1)\,2^{2s_\ell+1}\,\Gamma\!\left(s_\ell+\tfrac12\right)\Gamma\!\left(s_\ell+\tfrac32\right)}\,(ka)^{2s_\ell+1},
\qquad \ell\geq1,
\label{eq:lowfreq}
\end{equation}
while $\tan\eta_0\simeq-(ka)^3/3$. At $\alpha=1$ we recover the textbook values
$\eta_0\simeq-(ka)^3/3$ and $\eta_1\simeq(ka)^3/6$, and with them the Rayleigh
amplitude of a rigid sphere,
$f\simeq-\frac{k^2a^3}{3}\left(1-\frac32\cos\theta\right)$.

\begin{figure}[t]
\centering
\includegraphics[width=\linewidth]{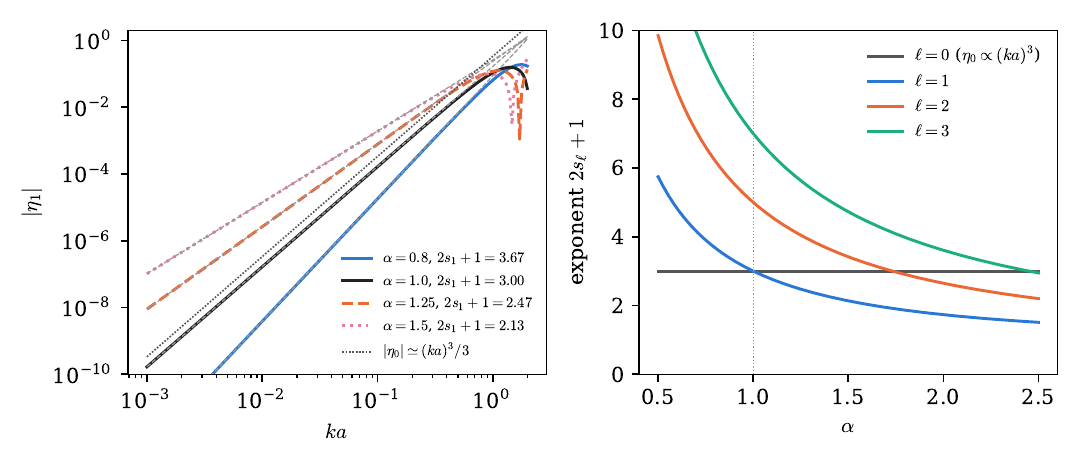}
\caption{Left: dipole core phase $|\eta_1|$ as a function of $ka$, with the
asymptotic law (\ref{eq:lowfreq}) shown as gray dashed lines and the monopole
phase $|\eta_0|\simeq(ka)^3/3$ as a dotted line. Right: low-frequency exponents
$2s_\ell+1$ as functions of $\alpha$. A multipole whose exponent lies below $3$
dominates the monopole at low frequency.}
\label{fig:lowfreq}
\end{figure}

The interesting change is in the dipole. Its exponent is now
\begin{equation}
2s_1+1=\frac{\sqrt{8+\alpha^2}}{\alpha},
\label{eq:s1}
\end{equation}
which falls below the monopole value $3$ exactly when $\alpha>1$. For an excess
of solid angle, low-frequency scattering by the core is therefore dominated by
the dipole and is strongly enhanced; for a deficit, the monopole takes over and
the scattering becomes nearly isotropic. The scattered intensity follows a
generalized Rayleigh law, $|f_{\rm core}|^2\propto a^2(ka)^{4s_1}$ for
$\alpha>1$, in place of the familiar $(ka)^4$. More generally,
$2s_\ell+1=\sqrt{4\ell(\ell+1)+\alpha^2}/\alpha$ is smaller than $3$ whenever
$\alpha^2>\ell(\ell+1)/2$, so the quadrupole also overtakes the monopole for
$\alpha>\sqrt3$, and higher multipoles follow at larger $\alpha$. The ordering
among the multipoles with $\ell\ge1$ never changes; only the monopole is
displaced (Fig.~\ref{fig:lowfreq}).

These statements refer to $f_{\rm core}$. At low frequency $f_{\rm top}=O(1/k)$
dominates $|f|$ everywhere except near the forward direction. The near-core
field, which is built from the same orders $s_\ell$ and phases $\delta_\ell$,
is what determines the radiation force in Sec.~\ref{sec:force}.

\subsection{Numerical results}
\label{sec:numerics}

All curves were computed from Eqs.~(\ref{eq:fcore}) and
(\ref{eq:ftop_split})--(\ref{eq:remainder}), using Bessel functions of real
order for $j_s$ and $y_s$ and the numerically stable form
$e^{2i\eta_\ell}=(y'_s+ij'_s)/(y'_s-ij'_s)$. We checked the code in three ways.
At $\alpha=1$ and $ka\le0.05$ it reproduces the Rayleigh amplitude to
$5\times10^{-3}$ and Eq.~(\ref{eq:lowfreq}) to $10^{-6}$. The closed forms $g$
and $h$ agree with the Abel-damped series to machine precision. Finally,
$f_{\rm top}$ obtained from Eq.~(\ref{eq:ftop_split}) agrees with a brute-force
Abel sum, extrapolated to zero damping, to better than $10^{-4}$ away from
$\theta_0$. The remainder $\mathcal F$ is already accurate to $10^{-6}$ with
$\ell_{\max}=2000$; the figures use $\ell_{\max}=6000$, and the truncation of
$f_{\rm core}$ is described in Appendix~\ref{app:numerics}. At no point is a
divergent series simply cut off.

\begin{figure}[t]
\centering
\includegraphics[width=\linewidth]{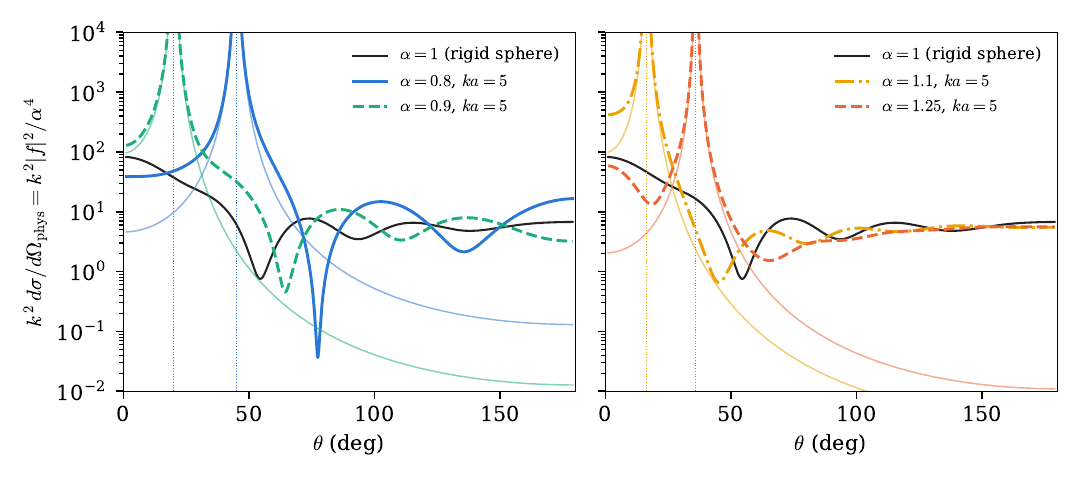}
\caption{Differential cross section per unit physical solid angle,
$k^2\,d\sigma/d\Omega_{\rm phys}=k^2|f|^2/\alpha^4$, at $ka=5$, for $\alpha<1$
(left) and $\alpha>1$ (right). Thick lines give the total amplitude
$f_{\rm top}+f_{\rm core}$ and thin lines the point-defect limit $f_{\rm top}$
($a\to0$); the black curve is the rigid sphere without the defect. Dotted
vertical lines mark $\theta_0=\pi|1-1/\alpha|$.}
\label{fig:dcs}
\end{figure}

\begin{figure}[t]
\centering
\includegraphics[width=\linewidth]{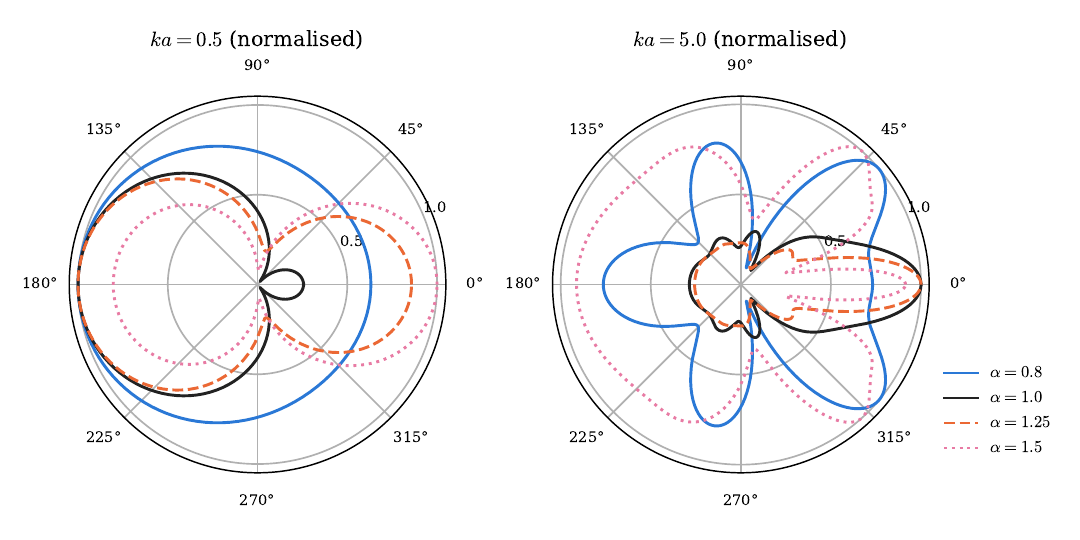}
\caption{Polar diagrams of the core form function $2|f_{\rm core}(\theta)|/a$,
each normalized to its maximum, for $ka=0.5$ (left) and $ka=5$ (right). The
incident wave travels from left to right.}
\label{fig:polar}
\end{figure}

Figure~\ref{fig:dcs} shows $d\sigma/d\Omega_{\rm phys}$ at $ka=5$. The forward
hemisphere is dominated by the singular ring at $\theta_0$, located exactly at
the deflection angle of Fig.~\ref{fig:rays}. At larger angles the dressed core
contribution takes over, and its interference pattern is visibly shifted with
respect to that of a rigid sphere without the defect. The core pattern alone is
shown in Fig.~\ref{fig:polar}. At $ka=0.5$ the backscattering lobe of a rigid
sphere in a homogeneous medium ($\alpha=1$) gives way, for $\alpha>1$, to a
forward dipolar lobe; this is the angular signature of the dipole dominance
discussed above. At $ka=5$ the defect redistributes the lobes and changes the
backscattered intensity considerably.

\section{Acoustic radiation force on the core}
\label{sec:force}

\subsection{Why the force must be computed on the core}

In a homogeneous fluid the radiation force on an obstacle is usually obtained
from the momentum flux through a large sphere surrounding it. The result is
$F=E_0\,\sigma_{\rm tr}$, where $E_0$ is the incident energy density and
$\sigma_{\rm tr}=(4\pi/k^2)\sum_\ell(\ell+1)\sin^2(\delta_\ell-\delta_{\ell+1})$
is the transport cross section~\cite{Westervelt,SilvaJASA}. The surface can be
moved to infinity because the time-averaged radiation stress (Brillouin) tensor,
\begin{equation}
\Pi^{ij}=\left(\frac{\langle p^2\rangle}{2\rho_0c_0^2}-\frac{\rho_0\langle v^2\rangle}{2}\right)g^{ij}+\rho_0\langle v^iv^j\rangle ,
\label{eq:brillouin}
\end{equation}
is divergence-free and flat space has translational Killing vectors.

In the medium (\ref{eq:metric}) the first property still holds,
$\nabla_j\Pi^{ij}=0$, but the second does not: the isometry group is only
$SO(3)$. The flux of $\Pi^{ij}\xi_j$ through a closed surface is independent of
the surface only when $\xi$ is a Killing vector, which here means a rotation.
Linear momentum is thus exchanged between the wave and the defect itself. The
far field tells the same story. Because
$\delta^{\rm top}_\ell-\delta^{\rm top}_{\ell+1}\to\pi\beta/2$, the series for
$\sigma_{\rm tr}$ diverges for every $\alpha\neq1$, so the far-field route to
the force is closed. The force on the core has to be computed on the core
surface, $r=a$.

There is a further subtlety. Without translational isometries, a
``displacement of the core along the axis'' has no intrinsic meaning in the
effective geometry, so the axial force must be defined. We take it to be the
flux of $\Pi^{ij}$ contracted with $\xi=\nabla(r\cos\theta)$ through $r=a$. This
field reduces to $\hat z$ for $\alpha=1$; for $\alpha\neq1$ it is neither a
Killing vector nor of unit norm. On a rigid surface $v_r=0$, only the isotropic
part of (\ref{eq:brillouin}) survives, and
\begin{equation}
F_z=-\oint_{r=a}\left(\frac{\langle p^2\rangle}{2\rho_0c_0^2}-\frac{\rho_0\langle v_\theta^2\rangle}{2}\right)\cos\theta\;\alpha^2a^2\,d\Omega ,
\qquad v_\theta=\frac{1}{\alpha a}\,\partial_\theta\phi ,
\label{eq:Fz}
\end{equation}
with $\alpha^2a^2d\Omega$ the physical area element and $v_\theta$ the physical
tangential velocity. In the metafluid of Sec.~\ref{sec:realisation} the
laboratory direction $\hat z$ is unambiguous, and the corresponding force,
which is closely related to $F_z$, is discussed in Sec.~\ref{sec:labframe}.

\subsection{Partial-wave formula}

With the asymptotic normalization of Sec.~\ref{sec:phases}, the total field is
\begin{equation}
\phi=\phi_0\sum_{\ell}(2\ell+1)\,i^\ell e^{i\delta_\ell}\left[\cos\eta_\ell\,j_{s_\ell}(kr)-\sin\eta_\ell\,y_{s_\ell}(kr)\right]P_\ell(\cos\theta).
\label{eq:totalfield}
\end{equation}
Its incoming spherical components coincide with those of a unit plane wave,
and, as shown in Sec.~\ref{sec:phases}, they carry the physical incident energy
density $E_{\rm inc}=\alpha^4E_0$, with $E_0=\rho_0k^2|\phi_0|^2/2$. On the
core, the Wronskian $j_sy_s'-j_s'y_s=x^{-2}$ puts the bracket in closed form,
\begin{equation}
W_\ell\equiv\cos\eta_\ell\,j_{s_\ell}(x)-\sin\eta_\ell\,y_{s_\ell}(x)=\frac{\cos\eta_\ell}{x^2\,y'_{s_\ell}(x)},\qquad x=ka .
\end{equation}
We insert (\ref{eq:totalfield}) into (\ref{eq:Fz}) and use
$\int_{-1}^{1}\mu P_\ell P_{\ell+1}d\mu=2(\ell+1)/[(2\ell+1)(2\ell+3)]$ together
with
$\int_{-1}^{1}\mu P^1_\ell P^1_{\ell+1}d\mu=2\ell(\ell+1)(\ell+2)/[(2\ell+1)(2\ell+3)]$.
The radiation-force function $Y\equiv F_z/(\pi a^2E_{\rm inc})$ then takes the
compact form
\begin{equation}
\boxed{\;Y(ka,\alpha)=-\frac{4}{\alpha^4}\sum_{\ell=0}^{\infty}(\ell+1)\,\sin\!\left(\delta_\ell-\delta_{\ell+1}\right)W_\ell W_{\ell+1}\left[\alpha^2-\frac{\ell(\ell+2)}{(ka)^2}\right]\;}
\label{eq:Y}
\end{equation}
The series converges quickly: once $\nu_\ell\simeq\ell/\alpha$ exceeds $ka$,
$W_\ell$ decays faster than any power of $\ell$. Note that $Y$ is linear, not
quadratic, in $\sin(\delta_\ell-\delta_{\ell+1})$, and therefore has no definite
sign.

We have checked Eq.~(\ref{eq:Y}) in two independent ways. For $\alpha=1$ it
coincides with $\sigma_{\rm tr}/\pi a^2$ to machine precision over the whole
range tested, $0.05\le ka\le8$, and it reproduces the Rayleigh limit of a rigid
sphere, $Y\simeq\frac{11}{9}(ka)^4$, in agreement with
$\int|f|^2(1-\cos\theta)\,d\Omega=11\pi k^4a^6/9$. For $\alpha\neq1$ we evaluated
(\ref{eq:Fz}) directly, building the fields from arbitrary-precision Bessel
functions and taking $\partial_\theta\phi$ by finite differences; the two
results agree to $10^{-9}$. In a homogeneous medium the near-field and
far-field expressions for the force are equal by momentum conservation. In the
monopole medium only the near-field one remains.

\subsection{Low-frequency limit: a pulling force}

For $x\ll1$ we have $W_\ell\simeq w(s_\ell)\,x^{s_\ell}$, with
$w(s)=\sqrt\pi/[(s+1)\Gamma(s+\frac12)2^{s}]$, and
$\delta_\ell\simeq\delta^{\rm top}_\ell$ for $\ell\ge1$. Two terms of
(\ref{eq:Y}) dominate: a potential (pressure) term from $\ell=0,1$ and a
kinetic term from $\ell=1,2$,
\begin{equation}
Y\simeq\underbrace{\frac{4}{\alpha^2}\sin\delta^{\rm top}_1\,w(s_1)\,(ka)^{s_1}}_{Y_p}
\;+\;\underbrace{\frac{24}{\alpha^4}\,\sin\!\left(\delta^{\rm top}_1-\delta^{\rm top}_2\right)w(s_1)w(s_2)\,(ka)^{s_1+s_2-2}}_{Y_k}.
\label{eq:Ylow}
\end{equation}
At $\alpha=1$ both sines vanish and we are back to the $(ka)^4$ law. For
$\alpha\neq1$, however, the force is of first order in the topological phase.
The defect gives the field near the core a standing-wave (gradient) component
that a progressive wave in a homogeneous medium simply does not have.

\begin{figure}[t]
\centering
\includegraphics[width=\linewidth]{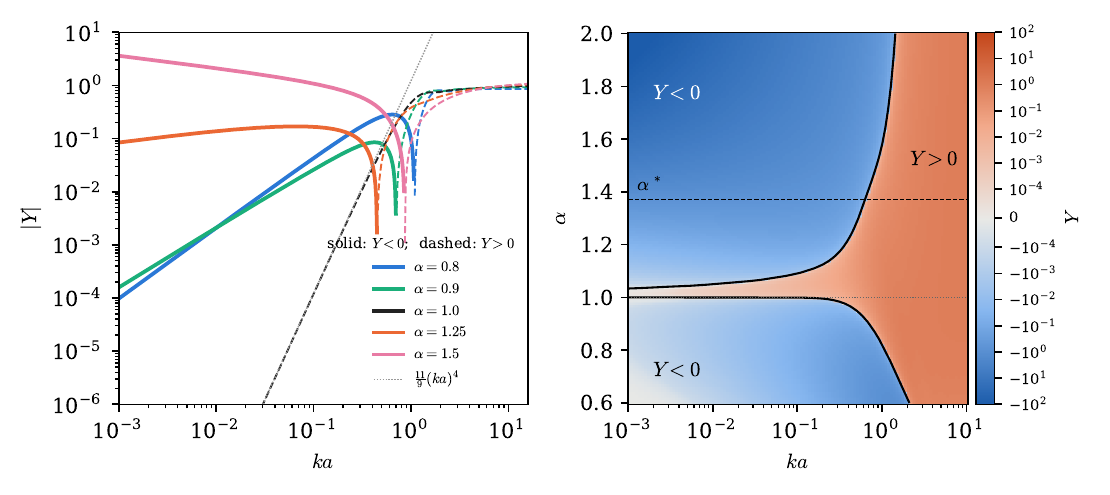}
\caption{Radiation-force function on the core, Eq.~(\ref{eq:Y}), normalized by
the physical incident energy density. Left: $|Y|$ as a function of $ka$, with
thick solid segments where $Y<0$ (pulling force) and thin dashed segments where
$Y>0$; the dotted line is the rigid-sphere Rayleigh law $\frac{11}{9}(ka)^4$.
Right: $Y$ in the $(ka,\alpha)$ plane on a symmetric logarithmic scale. The
solid contour is $Y=0$, and the dashed line marks $\alpha^\ast\simeq1.372$,
above which $|Y|\to\infty$ as $ka\to0$.}
\label{fig:force}
\end{figure}

The sign follows from Eq.~(\ref{eq:order}) and
$\delta^{\rm top}_1=\frac\pi2(1-s_1)$. For $\alpha<1$ we have $s_2>2$, so $Y_p$
is the leading term; when $1<s_1<3$, i.e.\ $1/\sqrt6<\alpha<1$, the phase lies
in $-\pi<\delta^{\rm top}_1<0$ and $Y_p<0$. For $\alpha>1$ instead $s_2<2$, the
kinetic term $Y_k$ wins, and since $s_2-s_1<1$ one has
$\delta^{\rm top}_1-\delta^{\rm top}_2<0$, hence $Y_k<0$. We conclude that,
\emph{for every $\alpha>1/\sqrt6\simeq0.41$ other than $\alpha=1$, the radiation
force pulls the core toward the source in a low-frequency band
$0<ka<ka_c(\alpha)$}. For stronger deficits, $\alpha<1/\sqrt6$, the
low-frequency sign is that of $\sin[\frac\pi2(1-s_1)]$ and alternates each time
$s_1$ crosses $3,5,\dots$. Numerically we find $ka_c\simeq1.09$, $0.71$, $0.45$
and $0.86$ for $\alpha=0.8$, $0.9$, $1.25$ and $1.5$, while $ka_c\to0$ as
$\alpha\to1$ (Fig.~\ref{fig:force}). On both sides of $\alpha=1$ the crossover
grows with $|\alpha-1|$, so the stronger the defect, the wider the pulling band.

A second threshold sits at $\alpha^\ast\simeq1.372$, where $s_1+s_2=2$. Beyond
it, $|Y|$ grows without bound as $ka\to0$, even though the force itself,
$F_z\propto a^{s_1+s_2}$, still goes to zero. The reason is that the velocity
near the center of a solid-angle-excess defect behaves as
$\nabla\phi\propto r^{s_\ell-1}$, which is singular when $s_1<1$; the rigid core
cuts this singularity off. For linear theory to remain valid one needs
$M_0(ka)^{s_1-1}\ll1$, where $M_0$ is the acoustic Mach number of the incident
wave, and this sets the lower limit of applicability of (\ref{eq:Ylow}).

Above $ka_c$ the force is positive and levels off close to the rigid-sphere
value $Y\to1$: at $ka=80$ we obtain $Y\simeq0.84$, $1.16$ and $1.22$ for
$\alpha=0.8$, $1.5$ and $2$. The relevant exponents and crossover frequencies
are collected in Table~\ref{tab:thresholds}.

\begin{table}[t]
\centering
\caption{Low-frequency exponents and the crossover frequency $ka_c$ below which
$Y<0$, obtained from Eq.~(\ref{eq:Y}). The last column gives $Y$ at $ka=20$.}
\label{tab:thresholds}
\begin{tabular}{ccccccc}
\hline
$\alpha$ & $s_1$ & $s_2$ & $2s_1+1$ & $s_1+s_2-2$ & $ka_c$ & $Y(ka=20)$\\
\hline
0.6 & 1.909 & 3.613 & 4.819 & 3.522 & 2.108 & 0.589\\
0.7 & 1.581 & 3.035 & 4.163 & 2.616 & 1.523 & 0.750\\
0.8 & 1.337 & 2.602 & 3.674 & 1.940 & 1.086 & 0.855\\
0.9 & 1.149 & 2.267 & 3.298 & 1.416 & 0.706 & 0.926\\
1.1 & 0.879 & 1.782 & 2.759 & 0.662 & 0.128 & 1.014\\
1.25 & 0.737 & 1.522 & 2.474 & 0.259 & 0.450 & 1.053\\
1.5 & 0.567 & 1.208 & 2.134 & $-0.225$ & 0.864 & 1.092\\
2.0 & 0.366 & 0.823 & 1.732 & $-0.811$ & 1.438 & 1.131\\
\hline
\end{tabular}
\end{table}

\subsection{Laboratory force in the metafluid realization}
\label{sec:labframe}

In the radially anisotropic fluid of Eq.~(\ref{eq:metafluid}) the laboratory
fields are the pressure $p$ and the velocity
$\mathbf v=(i\omega)^{-1}\hat\rho^{-1}\nabla p$. On the rigid core $v_r=0$, and
the tangential velocity is $v_t=(i\omega\rho_ta)^{-1}\partial_\theta p$.

Further analysis requires an additional assumption. We take the mean second-order excess
pressure on the core to be the Lagrangian density of the homogenized medium,
$\langle p^2\rangle/2\kappa-\rho_t\langle v_t^2\rangle/2$. This does not follow
from the linear effective parameters alone: the second-order response of a
metamaterial depends on its microstructure, and the Cartesian density tensor
$\rho_{ij}(\hat{\mathbf r})$ changes with direction. If the assumption holds,
then, using $\omega^2=k^2\kappa/\rho_r$ and $\rho_r/\rho_t=\alpha^{-2}$, the
integrand coincides term by term with that of Eq.~(\ref{eq:Fz}) for the same
field. The only difference is the area element, $a^2d\Omega$ in the laboratory
rather than $\alpha^2a^2d\Omega$, so that $F^{\rm lab}_z=F_z/\alpha^2$, where
$\hat z$ is now the genuine laboratory direction.

The sign of the force, the crossover $ka_c(\alpha)$ and the low-frequency
exponents are therefore unchanged in the laboratory. Seen from the laboratory,
the radial anisotropy singles out the center and breaks translation invariance,
and the momentum that is not delivered to the core is taken up by the structure
that sustains the anisotropy. A full-wave simulation of a layered metafluid,
with its microstructure resolved, would test both the assumption above and the
predicted sign reversal.

\section{Concluding remarks}
\label{sec:conclusions}

We have studied the scattering of sound by the rigid core of a defect with
global-monopole geometry, and the radiation force the wave exerts on it. The
analysis rests on three geometric facts. The deficit or excess of solid angle
persists at infinity; the curvature is $2(1-\alpha^2)/\alpha^2r^2$; and the
isometry group reduces to $SO(3)$. The first fact rules out a global plane wave
and leads to a phase-shift formulation in which the topological and core
contributions separate exactly. It is also responsible for the universal ray
deflection $\pi(1-1/\alpha)$, which shows up in the wave amplitude as a
singular ring. The second distinguishes the problem from that of the locally
flat cone. The third implies that the force on the core cannot be inferred from
the far field, whose transport cross section diverges, and has to be computed
on the core itself. The resulting physical behavior exhibits several distinctive features. At low frequency the core
scattering is reorganized: for $\alpha>1$ the dipole dominates and follows a
modified Rayleigh law $(ka)^{4s_1}$, and beyond $\alpha=\sqrt3$ the quadrupole
overtakes the monopole as well. The radiation force is of first order in the
topological phase, and for every $\alpha>1/\sqrt6$ other than $\alpha=1$ it is
negative below a crossover $ka_c(\alpha)$ that grows with $|\alpha-1|$. For
$\alpha>\alpha^\ast\simeq1.372$ the normalized force grows without bound as
$ka\to0$, as a consequence of the singular velocity field at the center of a
solid-angle-excess defect. This pulling force is produced by the medium alone,
with a plane-wave-like excitation and no shaping of the incident beam. In the
metafluid realization the laboratory force is $F_z/\alpha^2$, provided the
second-order stress of the metafluid is that of the homogenized medium, so the
same conclusions carry over.

Several extensions suggest themselves. A finite metafluid ball adds an outer
interface whose effect should be quantified. Elastic rather than rigid cores,
as well as viscous and thermal boundary layers, will modify the force at
finite frequency. In an elastic solid the full vector problem, including mode
conversion, remains to be solved. Off-center cores would probe the absence of
translational symmetry directly. Finally, finite-element simulations of the
anisotropic metafluid would provide an independent test of the predicted sign
reversal.

\section*{Acknowledgments}

C.F. acknowledges financial support from the Conselho Nacional de
Desenvolvimento Cient\'{\i}fico e Tecnol\'ogico (CNPq, Brazil)
under Research Productivity Grant No.~311781/2021-7.

\section*{Data availability}
The Python code used to compute the amplitudes, forces and figures is available
from the authors on request.

\appendix
\section{Small-argument forms and numerical scheme}
\label{app:numerics}

For $s>0$ and $x\to0$ one has
$j_s(x)\simeq\sqrt\pi\,x^s/[2^{s+1}\Gamma(s+\tfrac32)]$ and
$y_s(x)\simeq-2^s\Gamma(s+\tfrac12)/[\sqrt\pi\,x^{s+1}]$. Differentiating and
taking the ratio gives Eq.~(\ref{eq:lowfreq}), and the Wronskian
$j_sy'_s-j'_sy_s=x^{-2}$, valid for any real order, gives
$W_\ell\simeq w(s_\ell)x^{s_\ell}$. For $\ell=0$ the leading term of $j_0'$
vanishes, and one finds instead $\tan\eta_0\simeq-x^3/3$ and $W_0\simeq1$.

Spherical Bessel functions of real order were computed as
$j_s(x)=\sqrt{\pi/2x}\,J_{s+1/2}(x)$ and $y_s(x)=\sqrt{\pi/2x}\,Y_{s+1/2}(x)$.
We truncated the core series at
$\ell_{\max}=\max(\alpha,1)\,ka+10\,[\max(\alpha,1)\,ka]^{1/3}+25$ for
$f_{\rm core}$ and at $\max(\alpha,1)\,ka+15\,[\max(\alpha,1)\,ka]^{1/3}+30$
for $Y$; since $\nu_\ell\simeq\ell/\alpha$, this is enough for machine
precision. The remainder $\mathcal F$ of Eq.~(\ref{eq:remainder}) was summed up
to $\ell_{\max}=6000$, with a relative accuracy better than $10^{-6}$, and the
Legendre polynomials were generated from the three-term recurrence. The force
formula (\ref{eq:Y}) was checked against a direct evaluation of
Eq.~(\ref{eq:Fz}), with fields built from arbitrary-precision Bessel functions
and angular derivatives taken by finite differences; the two agree to
$10^{-9}$.


\end{document}